\documentclass[12pt,twoside]{article}
\usepackage{graphicx}

\begin{document}

\centerline{\bf On the existence of the ``classical trajectories''
of atoms in the Stern-Gerlach experiment}

\bigskip

\centerline{M. Dugi\' c$^{\ast}$\footnote{Department of physics,
Faculty of Science, Kragujevac, Serbia, Email: dugic@kg.ac.yu}, M.
Arsenijevi\' c$^{\ast}$, J. Jekni\' c-Dugi\' c$^{\ast\ast}$}

\bigskip

\centerline{\it $^{\ast}$Department of Physics, Faculty of
Science, Kragujevac, Serbia}

\centerline{\it $^{\ast\ast}$Department of Physics, Faculty of
Science, Ni\v s, Serbia}

\bigskip

PACS. 03.75.Dg- Atom and neutron interferometry

PACS. 03.65.Yz- Decoherence; open systems; quantum statistical
methods

03.65.Ta- Foundations of quantum mechanics; measurement theory

\bigskip

{\bf Abstract:} The widely accepted interpretation of the
Stern-Gerlach experiment  assumes the objective atomic
trajectories (the ``classical trajectories'') in front of the
screen. Following this interpretation, we perform an {\it ab
initio} analysis of the experiment and conclude that the objective
trajectories do not physically exist. The alternative to our
conclusion is substantially to change the model of the experiment.

\bigskip

{\bf 1. Introduction}

\bigskip

\noindent The Stern-Gerlach experiment is a paradigm of the
quantum measurement of spin [1]. However, there is still some
controversy in its physical interpretation. E.g., it is usually
assumed (probably due originally to Bohr, Pauli and Mott) that
the atoms in front of the screen bear the objective trajectories
(the ``classical trajectories'') [2, 3]. In this picture, the
atomic center of mass serves as the ``apparatus'' for the spin
measurement, due to the (classical) correlations between the
spin-projection (system $S$) and the (objectively existing,
semi-classical) trajectory of the atomic center of mass ($CM$).

Quantum mechanically, the state of the composite system $S+CM$
can be described by a ``mixed'' state $\hat \rho_{S+CM}$ e.g. of
the following form:

\begin{equation}
\hat\rho_{CM+S}=\frac{1}{2}\vert\uparrow\rangle_{S}\langle\uparrow\vert\otimes\vert{-}\rangle_{CM}\langle{-}\vert
+
\frac{1}{2}\vert\downarrow\rangle_{S}\langle\downarrow\vert\otimes\vert{+}\rangle_{CM}\langle{+}\vert
\end{equation}

\noindent where $\vert +\rangle_{CM}$ and $\vert -\rangle_{CM}$
are the states of the ``center-of-mass'' system (up and down
trajectories, respectively) and $\vert \uparrow\rangle_{S}$ and
$\vert \downarrow\rangle_{S}$ are the eigenstates of the spin
projection along the axis of the external magnetic field.

This scenario actually assumes the passive role of the screen
capturing the atoms. The screen is supposed passively to record
the atomic trajectories, which objectively exist {\it in front}
of the screen. However, this picture (scenario) is not the only
one possible in the context of the quantum mechanical formalism.
E.g., one may assume the active role of the screen, which actually
assumes the entangled state of the $S+CM$ system {\it in front}
of the screen  of e.g. the following (simplified) form (cf. Ref.
[4] for some details):

\begin{equation}
\vert\Psi\rangle_{CM+S}=\frac{1}{\sqrt{2}}\left(\vert\uparrow\rangle_{S}\vert{-}\rangle_{CM}
+\vert\downarrow\rangle_{S}\vert{+}\rangle_{CM}\right).
\end{equation}

The state eq. (2) elevates the active role of the screen: there
are not the objectively present trajectories in front of the
screen. Rather, the screen plays the active role in ``collapsing''
the state eq. (2) into the ``reduced'' state eq. (1), as
described in the von Neumann's quantum measurement theory [5].

The discrepancy between eq. (1) and eq. (2) is at the heart of the
modern quantum mechanics, quantum measurement theory [5, 6, 7],
the transition form quantum to classical [8, 9] as well as of the
quantum information/computation theory [10]. E.g., the second
scenario, as distinct from the first one, may imply some
usefulness of entanglement in eq. (2) as a quantum information
resource [4].

In order to solve this dilemma, we follow--as (in our opinion)
the easier option--the  interpretation formally described by eq.
(1). Our starting point is the standard model of SG experiment
[1], and we seek for a {\it proper extension} of the model that
might account for the ``objectively'' existing trajectories. More
precisely: we start an {\it ab initio} analysis of the atomic
system in search for the possible physical origin of the proposed
``objective trajectories''. We obtain the negative result: our
conclusion is that the "classical trajectories" can not be
considered to be physically realistic.

Given the $CM+S$ system is either in a mixed state eq. (1), or in
an entangled state eq. (2), we conclude, that in the SG
experiment, the {\it screen is responsible} for the appearance of
the classical information about the atom ``trajectory'' and,
consequently, about the spin projection. In other words: the
screen unfolds the (probably irreversible) retrieval of a
classical information from the quantum system.

In Section 2, we point out the necessity for the decoherence
effect in providing the ``classical trajectories'', and we go as
much as possible in pursuing this idea in the sense of searching
for the proper decoherence-based model that could  allow the
``classical trajectories''. Interestingly enough, the model(s)
obtained can not explain certain well-established experimental
finidings. Therefore, we are forced to consider the decoherence
model(s) as {\it physically unrealistic}. Section 3 is discussion.

\bigskip

{\bf 2. The decoherence model of SG experiment}

\bigskip

\noindent The standard model of SG experiment reads (cf. e.g.
[1]): a collimated beam of atoms (of the same chemical kind)
traverse the external magnetic field, which should be considered
as the external {\it classical} field {\it not} coupling with the
atomic degrees of freedom. The dynamics generated by the strong
magnetic field can be presented (approximately) by eq. (2), i.e.
by the existence of the entanglement in the system $CM+S$ [1].
Observing the definite paths on the screen reveals the
corresponding spin-projection of the atom.

However, bearing in mind eq. (2), the ``objective trajectories''
require an external action performed on the system--the action is
supposed to be responsible for the appearance of the reduced,
mixed state eq. (1). Needless to say, such an action should be of
the quantum-measurement type, and the quantum decoherence process
[2] seems {\it not to have any alternative} in this regard. In
other words: the (semi-classically) objective trajectories {\it
in front} of the screen require unfolding of the decoherence
process {\it in front of the screen}; then, the screen is
supposed passively to record the objectively existing
trajectories.  So, {\it our task reduces to searching for a
decoherence mechanism} that might justify physical objectivity of
the  ``classical trajectories''.

To this end, recently, a qualitative proposal has been made [11].
The core of this proposal reads: the SG experiment model
generally discards the so-called ``relative coordinates'' system
($R$) from consideration, and probably this system $R$ might play
the role of the (mesoscopic) {\it internal} environment for the
$CM$ system, i.e. to induce the decoherence of the $CM$'s
trajectories. The model stems an interesting and provoking
physical picture of the {\it internally} induced decoherence
[12]--which does not require any external environment. However,
this is fully a qualitative proposal that does not offer a
definite conclusion as to whether or not the corresponding model
(the decoherence-based model) of SG experiment can actually be
constructed. And this is the very issue of the present paper.

Generally, the center-of--mass ($CM$) system is defined by the
canonical transformations of the position-variables $\hat {\vec
r}_i$ of a system consisting of $K$ particles by:
\begin{equation}
\hat{\vec{R}}_{_{CM}}={\displaystyle\sum_{i=1}^K
m_i\hat{\vec{r}}_i}/{\displaystyle\sum_{i=1}^K m_i}.
\end{equation}

However, {\it simultaneously} and {\it unavoidably} are defined
the ``relative coordinates'' (that formally define the ``relative
system'' $R$) e.g. as:

\begin{equation}
\hat{\vec{\rho}}_{R\alpha}=\hat{\vec{r}}_i-\hat{\vec{r}}_j,
\alpha = 1, 2, ..., K-1.
\end{equation}

The system $R$ is generally discarded from the standard model of
SG experiment. And at this point appears the main idea of [11]:
probably the system $R$ might play the role of the environment
for $CM$ system--as the {\it missing link} to the ``classical
trajectories''.

Within the standard (and generally used) assumption that the SG
magnet is {\it not} a dynamical system but the source of the
external magnetic field for the atoms, it seems that the following
operational models do not  have any alternative. In other words:
the following models seem to exhaust the models that might fit
with the objective existence of the "classical trajectories".

\bigskip

{\bf 2.1 The atomic center of mass}

\bigskip

\noindent An atom is a collection of electrons ($E$), protons
($P$) and neutrons ($N$). Applying the transformations eqs. (3)
and (4) to the composite system $E+P+N$ introduces the
center-of-mass and the ``relative particles'' system for the whole
atom.

The "atom" can be sufficiently-well defined by the following
Hamiltonian:

\begin{equation}
\hat{H}=\sum_{i=1}^Z\hat{T}_{Ei}+\sum_{j=1}^Z\hat{T}_{pj}+\sum_{k=1}^{A-Z}\hat{T}_{nk}+\hat{V}^{ee}_{Coul}+
\hat{V}^{ep}_{Coul}+\hat{V}^{pp}_{Coul}+\hat{V}_{nucl}
\end{equation}

\noindent where $\hat{T}$ stands for the kinetic terms,
$\hat{V}_{Coul}$ for the Coulomb interaction
 of the pairs of particles ({\it ee}-the electrons, {\it ep}-the electron-proton,
{\it pp}-the protons pairs), and the nucleon interaction for a
pair $(n, n')$ of nucleons is given e.g. by [13]:
\begin{equation}
\hat{V}_{nucl}^{nn'}\equiv{-}\gamma^{2}\displaystyle\frac{\exp({-\mu|\hat{\vec{r}}_{n}-\hat{\vec{r}}_{n{'}}|})}{|\hat{\vec{r}}_{n}-\hat{\vec{r}}_{n{'}}|}
\end{equation}

\noindent where $\gamma$ is a constant and  r=$\frac{1}{\mu}$ is
the range of the nuclear forces. For simplicity, we omit the
comparatively weak interactions, such as the spin-spin or
spin-orbit interactions in the atom.

As apparent from eq. (5), the canonical transformations eqs. (3)
and (4) give for the atomic Hamiltonian: \noindent
\begin{equation} \hat H = \hat T_{CM} + \hat H_R + \hat H_{CM+S},
\end{equation}

\noindent where $\hat H_{CM+S} = \mu_B B(\hat z_{CM}) \otimes
\hat S_z$ is the standard term [1, 11] coupling the  center of
mass ($\hat z_{CM}$) and the atomic spin ($\hat S_z$), while the
$R$-system's self-Hamiltonian reads:

\begin{equation}
\hat H_R
=\sum_{\alpha=1}^{Z+A-1}\hat{T}_{R\alpha}+\hat{V}^{(R)}_{nucl}+\hat{V}^{(R)}_{Coul}+\hat{M}^{(R)}_{\eta\nu},
\end{equation}
where $Z$, $A$ are the atomic and the mass numbers, respectively,
and $\hat{M}^{(R)}_{\eta\nu}$ is the internal interaction in $R$
 [14].

Regarding eq. (7), it is important to note: being the
distance-dependent, all the original interactions (the Coulomb
interaction and the nuclear interaction  in eq. (5)) transform
into the ``external fields'' (the one-particle potentials $V(\hat
{\vec \rho}_{Ri})$) for the ``relative particles''  system
 $R$. These effective potentials are the terms of
 the $R$'s self-Hamiltonian $\hat H_R$.
 Needless to say, this gives the {\it exact separation} of (non-interaction between) $CM$ and $R$
 that does  not leave a room for the desired
 decoherence of the $CM$ states [2, 15].

In the terms of the quantum states, the initial state, e.g.,

\begin{equation}
\frac{1}{\sqrt{2}}(\vert \uparrow\rangle_S + \vert \downarrow
\rangle_S) \vert \Psi \rangle_{CM}|0 \rangle_{R}
\end{equation}

\noindent dynamically transforms as presented by the following
simplified expression:

\begin{equation}
\hat U \frac{1}{\sqrt{2}}(\vert \uparrow \rangle_S
 + \vert \downarrow\rangle_S)\vert \Psi \rangle_{CM}|0 \rangle_{R}
= \frac{1}{\sqrt{2}} (\vert \uparrow\rangle_S \vert -\rangle_{CM}
+ \vert \downarrow\rangle_S \vert +\rangle_{CM}) \vert 0
\rangle_R,
\end{equation}

\noindent where $\hat U$ is generated by $\hat H$ eq. (7). After
``tracing out'' the environment $R$, one obtains the entangled
state eq. (2)--there are not the ``classical trajectories'', which
require decoherence, i.e. the interaction in the $CM+R$ system.

 However, in order to make our search for the desired
 interaction complete, we move a step further as presented in the next
 section.

\bigskip

{\bf 2.2 The atomic-nucleus center of mass}

\bigskip

\noindent More than 99.99 per-cents of the atomic mass is placed
in the atomic nucleus. Practically, it is truly hard to
distinguish between the atomic and the nucleus center-of-mass
systems. So, we investigate another application of eqs. (3), (4):
we introduce the center-of-mass system and the ``relative
system'' for the atomic {\it nucleus} while leaving the electrons
variables intact.

Introducing the collective degrees of freedom of the atomic
nucleus is the standard procedure in nuclear physics [16]. On the
other side, the similar idea appears in certain models of the
quantum measurement theory, unfortunately not yet being fully
elaborated [17]. So, introducing the center of mass of the atomic
nucleus not yet involving the electrons is physically legitimate
a procedure.

Then, ``atom'' is a composite system defined as $E + CM + R + S$,
where $E$ stands for the electrons-system, $CM$ and $R$ for the
nucleus center-of-mass and the ``relative'' systems,
respectively,  while $S$ is the atomic spin.

Now, the standard model of the SG experiment is defined by the
following form of the atomic Hamiltonian (in analogy with eq.
(7)):

\begin{equation}
\hat H = \hat H_E + \hat T_{CM} + \hat H_R + \hat H_{CM+S} + \hat
H_{E+CM+R}.
\end{equation}

\noindent Certainly, the Hamiltonian $\hat H$ in eq. (11) and eq.
(7) is the one and the same observable--it is just written in the
different forms, yet in eq. (11) appearing the interaction term
for $E$, $CM$ and $R$ systems:

\begin{equation}
\hat{H}_{E+CM+R}=k\displaystyle\sum_{i=1}^Z\displaystyle\sum_{j=1}^Z\displaystyle\frac{1}{|\hat{\vec{r}}_{Ei}-\hat{\vec{R}}_{CM}-\displaystyle\sum_{\alpha=1}^{A-1}\omega^{(j)}_{\alpha}{\hat{\vec{\rho}}}_{R\alpha}\!\!\!^{\scriptstyle{(j)}}|}
,
\end{equation}

\noindent where
$\hat{\vec{R}}_{CM}+\displaystyle\sum_{\alpha=1}^{A-1}\omega^{(j)}_{\alpha}{\hat{\vec{\rho}}}_{R\alpha}\!\!\!^{\scriptstyle{(j)}}=
\hat {\vec r}_{pj}$, and $\hat {\vec r}_{pj}$ represents the
$j$-th proton position. So, the tripartite interaction $\hat
H_{E+CM+R}$ is a particular form of the Coulomb interaction
between the atomic electrons and the protons. Interestingly
enough, this tripartite interaction can be reduced to a bipartite
interaction coupling $CM$ and $R$ systems as follows.

The close inspection of the rhs of eq. (11) justifies the
 application of the {\it adiabatic approximation} that
in its zeroth order separates the electrons system from the rest.
More precisely (cf. Appendix 1): the electrons are too light
relative to both the $CM$- and $R$-mass, thus allowing the
standard procedure of the adiabatic approximation [18, 19, 20]. On
the
 other side, for the realistic atoms (not too large $Z$), the $CM$ and $R$
mass-ratio does not allow the application of the adiabatic
 approximation. So, we expect the approximate separation of
the electrons-state from the rest, $CM + R + S$,  as well as
non-negligible entanglement between $CM$ and $R$. Formally,
 the state now reads:

\begin{equation}
\vert \chi \rangle_E \vert \Phi\rangle_{CM+R+S} + \vert
O(\kappa)\rangle_{E+CM+R+S},
\end{equation}

\noindent where the small term (that bears entanglement, in
general, of all of the subsystems) is of the norm $\kappa^{3/4}$,
where $\kappa = max \{\kappa_1, \kappa_2\}$, and $\kappa_i$ are
the corresponding mass ratios, cf. Appendix 1.

In order to obtain the dynamics of the ``slow'' system $CM + R$
(i.e. of $CM + R + S$), one should discard the electrons system
as (cf. Appendix 1):

\begin{equation}
\hat H_{CM+R+S} \equiv _E\langle \chi \vert \hat H \vert \chi
\rangle_E \cong \hat T_{CM} + \hat H_R + \hat H_{CM+S} + \hat
H_{CM+R},
\end{equation}

\noindent where

\begin{equation}
\hat H_{CM+R} \equiv _E\langle \chi \vert \hat H_{E+CM+R} \vert
\chi\rangle_E
\end{equation}
represents the {\it effective} (the electrons--system {\it
mediated}) interaction between $CM$ and $R$.

Now, due to the two interaction terms, $\hat H_{CM+S}$ and $\hat
H_{CM+R}$ in eq. (14), it is straightforward dynamically to obtain
entanglement in the dominant term of the state in eq. (13), $\vert
\Phi\rangle_{CM+R+S}$. Actually, for the initial state eq. (9)
and in analogy with eq. (10) one obtains:

\begin{equation}
\hat U \frac{1}{\sqrt{2}}(\vert \uparrow\rangle_S + \vert
\downarrow \rangle_S) \vert \Psi \rangle_{CM} \vert 0 \rangle_{R}
\cong\frac{1}{\sqrt{2}} (\vert \uparrow\rangle_S \vert -
\rangle_{CM} \vert 1\rangle_R + \vert \downarrow\rangle_S \vert +
\rangle_{CM} \vert 2\rangle_R)
\end{equation}

\noindent where $\hat{U}$ is generated by $\hat{H}$ represented in
eq. (14). Now, assuming the orthogonality
$_R\langle1|2\rangle_R\approx0$, by tracing out the
``environment'' $R$ from the  rhs of eq. (16) follows the mixed
state eq. (1) for $CM+S$ system, {\it as desired}.

The interaction $\hat H_{CM+R}$ is analyzed in detail in [21] and
is briefly presented in Appendix 2. This interaction provides the
``minimal uncertainty states'' as the good (approximate) pointer
basis--in agreement with the standard model of SG experiment
[1]--and for the larger atoms ($Z \sim 10$), the interaction
scales approximately as $Z^2$.

\bigskip

{\bf 2.3 Inconsistency of the decoherence-based model with certain
experiments}

\bigskip

\noindent The model of Section 2.2 bears certain straightforward
consequences. Here, we give only those of importance for our
conclusion; for more detailed discussion see Ref. [21].

First, as obvious from eq. (12), without $R$, there is not the
basis for the decoherence effect. So, for the particles not
bearing $R$ (e.g. electrons, neutrons, protons), likewise for the
hydrogen atom, for which $A = 1$ and $R$ simply does not exist,
one can not expect the occurrence of decoherence. Second,
existence of $R$ does not suffice for the occurrence of
decoherence. According to eq. (12), the presence of the electrons
is necessary in order to obtain the effective (the
electrons-mediated) interaction $\hat H_{CM+R}$; otherwise, the
model  reduces to the scenario of Section 2.1. So, the model
proposes nonappearance of the SG effect also for the bare atomic
nuclei.

While certain predictions of the model of Section 2.2 fit with
some experimental findings, there is a couple of the experimental
observations/results being yet in some inconsistency (and
probably in contradiction) with the model.

First, the model of Section 2.2 predicts the absence of the SG
effect for the hydrogen atom--in contradiction with the well-known
experiments performed first by Phipps and Taylor [22]. Second,
this
 model does not allow a room for  explanation of
certain atomic interference experiments [23, 24].

As to the later, it is worth stressing: the existence of the
environment $R$ i.e. of the decoherence of the center-of-mass
trajectories, makes some interference-procedures for the atoms
impossible. Actually, in the notation of Section 2.2, the
standard equality of ``quantum erasure'' reads [24]:

\begin{equation}
2^{-1/2} (\vert \uparrow\rangle_S \vert -\rangle_{CM} + \vert
\downarrow \rangle_S \vert +\rangle_{CM}) = 2^{-1} [\vert
\rightarrow\rangle_S (\vert -\rangle_{CM} + \vert + \rangle_{CM})
+ \vert \leftarrow\rangle_S (\vert -\rangle_{CM} - \vert
+\rangle_{CM})],
\end{equation}

\noindent where appear the (spin $x$-projection) $\hat S_x$
eigenstates on the rhs of eq. (17), $\vert \rightarrow \rangle_S$
and $\vert \leftarrow \rangle_S$. Of course, the measurement of
$\hat S_x$ can give the value $+1$ with the probability $1/2$ and
with the final state:

\begin{equation}
2^{-1/2} \vert \rightarrow\rangle_S (\vert -\rangle_{CM} + \vert +
\rangle_{CM})
\end{equation}

\noindent providing the {\it reunion} (interference) of the
initial coherence of the different trajectories--as
experimentally verified [23, 24].

The presence of the environment $R$ does not allow the
trajectories-reunion by the quantum measurement of $\hat S_x$.
Actually, the inclusion of  $R$ gives the rhs of eq. (10) and the
measurement of $\hat S_x$ gives rise to  the final state:

\begin{equation}
2^{-1/2} \vert \rightarrow\rangle_S (\vert -\rangle_{CM} \vert
1\rangle_R + \vert +\rangle_{CM} \vert 2\rangle_R)
\end{equation}

\noindent i.e. to entanglement in the $CM+R$ system. Effectively,
the $CM$ system is in the mixed state $1/2(\vert -\rangle_{CM}
\langle - \vert + \vert + \rangle_{CM} \langle + \vert)$, instead
of the coherent state eq. (18).

\bigskip

{\bf 3. Discussion}

\bigskip

\noindent If the magnetic field is not a dynamical system {\it
and} the ``center-of-mass''  should bring the information about
the atomic spin, then we {\it do not see any alternative} to our
conclusion that the ``classical trajectories''-based
interpretation of SG experiment should be refuted.

So, it is of interest to answer the following question: which
assumptions about the experiment could question our conclusion.
The following list in this regard is of interest. Actually, (i)
one may assume that the decoherence process is not of interest,
e.g. that there exists an alternative to the decoherence process
in providing the ``classical trajectories''. On the other side, if
decoherence {\it can not be circumvented}, then one may (ii)
assume that the magnetic field plays a role of the environment,
or (iii) that some external, not yet recognized environment is
effective, or (iv) that another internal environment should be
recognized. One may also speculate (v) that the center-of-mass is
not of interest (e.g. the screen monitors another collective
observable of the atom). In answer to these remarks,
respectively, we want to emphasize: (i) to the best of our
knowledge, the decoherence effect is currently the only candidate
for providing the (approximately) classical behaviour of a
genuinely (yet open) quantum system, i.e. to ``produce'' the
``classical trajectories'' [2]. Bearing this in mind, (ii)
considering the magnetic field as a dynamical system in the
SGE-like situations could be in contradiction e.g. with the
neutron interferometry experiments [25]. The points (iii)-(v)
seem virtually intractable to us  as  requiring a substantial
reconsideration/modeling of SG experiment from the very beginning.
So, it is fair to say, that, as yet, we do not see any reasonable
alternative to our conclusion on nonexistence of the ``classical
trajectories'' in the Stern-Gerlach experiment.

As long as we adopt the standard model of SG experiment, that
assumes the magnet (i.e. the magnetic field) not to represent a
dynamical system, we consider our analysis to be complete and
therefore {\it conclusive}. Our logic is as follows. Assuming that
there is not any external environment in SG experiment, we are
{\it forced} to look for another environment among the {\it
internal} degrees of freedom. Of course, of interest is the
center-of-mass system, and the application of the canonical
transformations eqs. (3), (4) seem essentially to be {\it without
alternative}. Certainly, there exist the alternatives to the
definition eq. (4) of the ``relative coordinates'' and therefore
the different formal definitions of the ``relative system'' $R$.
Nevertheless, and this is the point, existence of the (no matter
how formally defined) system $R$ is unavoidable. So, the system
$R$ is the {\it only candidate} for playing the role of the
$CM$'s (internal) environment. Now, due to nonexistence of the
interaction between the {\it atomic} subsystems $CM$ and $R$,
there does {\it not seem to appear any alternative} to the model
described in Section 2.2. Finally, as described in Section 2.3,
the model fails to describe certain well-established experimental
findings. E.g., as long as the trajectories reunion is performed
by measuring the proper observable of $S$, and not of a composite
system ($S+R$) observable, the effect eq. (18) can not  {\it in
principle} be obtained.

While the different formal definitions of the ``relative
coordinates'' i.e. of the system $R$ are possible, the variations
in this regard seem nothing to change in our conclusion: by
discarding the decoherence effect as a physical basis of the
"classical trajectories", we promote the screen as the ``quantum
apparatus'' responsible for acquiring a classical information
from the atoms impinging on the screen in SG experiment.

Of course, this does not mean that the decoherence model of
Section 2.2 is formally wrong, or that it generally gives the
wrong predictions. The model is derived from the first principles
and nicely reproduces certain well-known experimental findings
(cf. Ref. [21] for some details). The model is just in
inconsistency with certain experimental findings as emphasized
above--that is the reason we are forced to consider the model
{\it not to be physically realistic}. The physical reasons for
this might be [21]: (a) that the screen observes the atomic- not
yet the nucleus- center-of-mass--in agreement with the model of
Section 2.1), or (b) that the (formally possible) decoherence due
to $\hat H_{CM+R}$ eq. (15) is not physically efficient due to the
fact that $R$ is a small environment. In any case, we conclude
that the "classical trajectories"-based interpretation of the
Stern-Gerlach experiment should be refuted--which is our
conclusion.

\bigskip

{\bf Acknowledgements} The work on this paper is financially
supported by Ministry of Science Serbia, under contract no 141016.

\bigskip

{\bf References}

\bigskip
\begin{flushleft}
[1] D. Bohm, ``Quantum theory'', Prentice Hall,Inc.,  New York,
1951

[2] W. Pauli, in {\it Handbuck der Physik}, Ed.  S. Flugge
(Springer, Berlin, 1985), p. 165

[3] H. Batelaan, T. J. Gay, J. J. Schwendimann, {\it Phys. Rev.
Lett.} {\bf 79}, 4517 (1997)

[4] T. R. Oliveira, A. O. Caldeira, ``Coherence and Entanglement
    in a Stern-Gerlach experiment'', eprint arXiv quant-ph/0608192v1
    24 Aug 2006

[5] J. von Neumann, ``Mathematical Foundations of Quantum
Mechanics'',
    Princeton University Press, Princeton, 1955

[6] {\it Quantum Theory and Measurement}, Eds. J. A. Wheeler and
W. H. Zurek, (Princeton Univ. Press, Princeton, 1983)

[7] P. Grigolini, ``Quantum Mechanical Irreversibility and
Measurement'', World Scientific, Singapore, 1993

[8] D. Giulini, E. Joos, C. Kiefer, J. Kupsch, I.-O. Stamatescu
    and H.D. Zeh, ``Decoherence and the Appearance of a Classical
    World in Quantum Theory'', Springer, Berlin, 1996

[9] H. D. Zeh, ``The Physical Basis of The Direction of Time'',
Springer-Verlag, 3rd edition, Berlin, 1999

[10] M. Nielsen and I. Chuang, ``Quantum Computation and Quantum
Information'', Cambridge, UK, 2000

[11] M. Dugi\' c, Europ. Phys. J, D {\bf 29}, 173 (2004)

[12] R. Omn$\grave{e}$s, ``The Interpretation of Quantum
Mechanics'',
     Princeton University Press, Princeton, 1994

[13] B. L .Cohen, ``Concepts of Nuclear Physics'', McGrawHill Book
Company, New York, 1971

[14] R. McWeeny, ``Methods of Molecular Quantum Mechanics'',
     Academic Press, New York, 1978

[15] M. Dugi\' c, Physica Scripta {\bf 53}, 9 (1996)

[17] H. D. Zeh, ``Roots and Fruits of Decoherence'',  Eprint
arXiv: quant-ph/0512078 v1 10 Dec 2005

[18] A. Messiah, ``Quantum Mechanics'', North--Holand Publishing
    Company, Amsterdam, 1976

[19] P. Atkins, R. Friedman, ``Molecular Quantum Mechanics'',
Oxford Univ. Press, Oxford, 2005

[20] L. A. Gribov, S. P. Mushtakova, ``Quantum Chemistry'',
Gardariki, Moscow, 1999 (in Russian)

[21] M. Dugi\' c, M. Arsenijevi\'c,  J. Jekni\' c-Dugi\'c, ``The
internal-environment model of the Stern-Gerlach experiment'',
Eprint arXiv: quant-ph/0809.4376v1 v1 25 sep 2008

[22] T. E. Phipps and J. B. Taylor, Phys. Rev. 29, 309 (1927)

[23] A. D. Cronin, J. Schmiedmayer, D. E. Pritchard, {\it Atoms
Interferometers}, eprint arXiv quant-ph/0712.3703

[24] S. D\"urr, G. Rempe, {\it Optics Communications} {\bf 179},
323 (2000)

[25] H. Rauch, S. A. Werner, {\it Neutron Interferometry}, Oxford
Univ. Press, Oxford, 2000

\end{flushleft}

\bigskip

{\bf Appendix 1}

\bigskip

\noindent Let us focus on the atomic nucleus with the
simplification of the equal masses of the protons and the
neutrons, $m$. Then, eqs. (3), (4) define the total (the $CM$
system) mass $M = Am$ and the ``relative mass'' $\mu$ (for all the
``relative particles'' enumerated by $\alpha$ in eq. (4)) as:

\begin{equation}
\mu = (1 - A^{-1}) m.
\end{equation}

The kinetic terms for the electrons, the $CM$ system and the $R$
system, as implicit in eq. (11) read, respectively, as follows:

\begin{equation}
\hat{T}_{E}=\displaystyle\frac{\hat{\vec{P}}^2_{E}}{2m_E},\qquad
\hat{T}_{CM}=\displaystyle\frac{\hat{\vec{P^2}}_{CM}}{2M},\qquad
\hat{T}_{R\alpha}=\displaystyle\frac{\hat{\vec{P^{2}}}_{R\alpha}}{2\mu_{\alpha}},
\forall{\alpha = 1, 2, ..., K - 1}.
\end{equation}

With eqs. (20), (21) in mind, there appear the three   parameters,
\begin{equation}
\kappa_1\equiv\frac{m_e}{M},\qquad
\kappa_2\equiv\frac{m_e}{\mu},\qquad \kappa_3\equiv\frac{\mu}{M},
\end{equation}

\noindent that allow the standard adiabatic-approximation
considerations [18, 19, 20]. For the {\it realistic atoms},
$Z\stackrel{<}{\sim}10^{2}$, one may state the following
estimates:

\begin{equation}
\kappa_1\stackrel{<}\sim {10^{-4}},\qquad
\kappa_2\stackrel{<}\sim {10^{-3}},\qquad
\kappa_3\stackrel{>}\sim {10^{-2}}.
\end{equation}

Then the values eq. (23) justify the applicability of the {\it
adiabatic approximation} [2, 13, 14] for $E+CM+R$ as follows: the
small values of $\kappa_{1,2}$ justify the adiabatic cut of the
electronic system $(E)$ from both $CM$ and $R$ systems, while
$CM$ and $R$ can not be properly mutually separated.

Now, the standard adiabatic approximation stems [18, 19, 20]: (a)
the exact state of $E+CM+R+S$ system reads

\begin{equation}
\vert \chi \rangle_E \vert \Phi\rangle_{CM+R+S} + \vert
O(\kappa)\rangle_{E+CM+R+S},
\end{equation}

\noindent where $\kappa=max\{\kappa_1,\kappa_2\}$, while (b) the
``slow'' system $CM+R+S$ is described by the following effective
Hamiltonian

\begin{equation}
\hat{H}_{CM+R+S}\cong_E\langle\chi|\hat{H}|\chi\rangle_E.
\end{equation}

\bigskip

{\bf Appendix 2}

\bigskip

\noindent For $|\chi\rangle_E$ in eq. (15), we take the
$Z$-electrons Slater determinant constructed from the
hydrogen-atom states. This simplification ease our task yet
without introducing a significant quantitative error. Then,
formally, our task reduces to calculating the following
expression [21]:

\begin{equation}
\hat{H}_{CM+R}=kZ\displaystyle\sum_{i=1}^Z\int\displaystyle\frac{|\phi_i(\vec{\xi})|^2}
{|{\vec{\xi}}-\hat{\vec{\Omega}}_{CM+R}|}d^{3}\vec{\xi}.
\end{equation}

\noindent where
$\hat{\vec{\Omega}}_{CM+R}\equiv-\vec{r}_{{CM}}\hat{I}_E+\hat{\vec{R}}_{CM}+\displaystyle\sum_{\alpha=1}^{A-1}\omega_{\alpha}\hat{\vec{\rho}}_{R{\alpha}}$.
Taking the point-like  nucleus gives rise to the shift $
\hat{\vec{r}}_{Ei}\longrightarrow \hat {\vec \xi}_{Ei} =
\hat{\vec{r}}_{Ei}-\vec{r}_{CM}\hat{I}_E$ as explicit in eq. (26).

The details of calculating the rhs of eq. (26) are given in Ref.
[21], and the result for the atoms with the ``closed shells''
reads:

\begin{eqnarray}
\hat{H}_{CM+R}&=\nonumber&kZ\sum_{n}\sum_{\ell=0}^{n-1}\sum_{g=0}^{n-\ell-1}\sum_{t=0}^{2g}\frac{2\ell+1}{2n
2^{2(n-\ell-1)}}{{2(n-\ell-1)-2g}\choose{n-\ell-1-g}}\times
\nonumber \\&&
\times\frac{(2g){!}}{g{!}(2\ell+1+g){!}}{{2g+2(2\ell+1)}\choose{2g-t}}\frac{(-2)^t}{t{!}}\nonumber\\&&\Bigg\{(2\ell+t+2){!}
\Bigg(1-\exp\bigg(-\frac{2Z\hat{\Omega}}{na_{\mu}}\bigg)\sum_{f=0}^{2\ell+t+2}\frac{(\frac{2Z\hat{\Omega}}{na_{\mu}})^{f}}{f{!}}\Bigg)\hat{\Omega}^{-1}\nonumber\\&&+
\frac{2Z}{na_{\mu}}(2\ell+t+1){!}\exp\bigg(-\frac{2Z\hat{\Omega}}{na_{\mu}}\bigg)\sum_{f=0}^{2\ell+t+1}\frac{(\frac{2Z\hat{\Omega}}{na_{\mu}})^{f}}{f{!}}\Bigg\}.
\end{eqnarray}

\noindent The notation is as follows: the big brackets indicate,
as usual, the binomial coefficients and the sign "$!$" stands for
the factorial.

For the atoms for which $Z \sim 10$, one can simplify eq. (27) and
to estimate that the interaction scales as $Z^2$. Finally, as it
can be easily shown, the minimal-uncertainty states (the
``coherent states'') appear as the approximate ``pointer basis''
[2, 15] for the model eq. (27)--in accordance with the standard
model of SG experiment [1].

\bigskip

\end{document}